\documentstyle[aps,epsfig,multicol]{revtex}
\def\breakon{\end{multicols}\widetext\vspace{-.2cm}
\noindent\rule{.48\linewidth}{.3mm}\rule{.3mm}{.3cm}\vspace{.0cm}}

\def\breakoff{\vspace{-.2cm}
\noindent
\rule{.52\linewidth}{.0mm}\rule[-.27cm]{.3mm}{.3cm}\rule{.48\linewidth}{.3mm}
\vspace{-.3cm}
\begin{multicols}{2}
\narrowtext}

\renewcommand{\phi}{\varphi}

\newcommand{\be}{\begin{equation}}
\newcommand{\ee}{\end{equation}}
\newcommand{\bea}{\begin{eqnarray}}
\newcommand{\eea}{\end{eqnarray}} 
\newcommand{\HH}{{\cal H}}

\newcommand{\tr}{{\rm tr\/}\,}
\newcommand{\p}{\partial}
\newcommand{\s}{\sigma}

\newcommand{\lb}{\left[}
\newcommand{\rb}{\right]}
\newcommand{\lp}{\left(}
\newcommand{\rp}{\right)}

\begin{document}

\title{Fermi-Edge Resonance and Tunneling in 
Nonequilibrium Electron Gas}
\author{D. A. Abanin and L. S. Levitov}
\address{
Department of Physics,
Center for Materials Sciences \& Engineering, \\
Massachusetts Institute of Technology, 
77 Massachusetts Avenue,
Cambridge, MA 02139}

\maketitle
\begin{abstract}
Fermi-edge singularity changes in a dramatic way 
in a nonequilibrium system, acquiring features
which reflect the structure of energy distribution.
In particular, it splits into several components if the energy 
distribution exhibits multiple steps. 
While conventional approaches, such as bosonization,
fail to describe the nonequilibrium problem,
an exact solution for a generic energy distribution
can be obtained with the help 
of the method of functional determinants.
In the case of a split Fermi distribution,
while the `open loop' contribution to Green's function
has power law singularities,
the tunneling density of states profile
exhibits broadened
peaks centered at 
Fermi sub-levels.
\end{abstract}


\vspace{-10mm}
\begin{multicols}{2}

\narrowtext
Properties of quantum systems
can change drastically when they are driven out of equilibrium. 
This is especially true for transport 
in nanodevices\cite{review}, such as quantum dots and quantum wires, 
where energy relaxation takes place outside the device.
Transport in interacting systems is often difficult to describe
by the methods developed for analyzing 
equilibrium\cite{Kaminski,Moore,Coleman,Konik0102,Paaske,Komnik04},
which makes exact solutions 
outside equilibrium scarce and valuable. 

Fermi-edge singularity\cite{Mahan,Nozieres_deDominicis} (FES)
is a dramatic manifestation of 
interactions and correlations in electron liquid.
It can be observed
in a resonant tunneling experiment\cite{MatveevLarkin,Geim}
as a power law resonance which peaks at the Fermi level.
Being one of the few exactly solvable problems
describing transport in strongly
interacting systems, FES has been thoroughly explored
in a variety of situations, including 
quantum wires\cite{Ogawa92,Prokof'ev,Komnik97}, 
quantum Hall edge states\cite{Affleck94}
and quantum dots\cite{Bascones}.
However, apart from recent work\cite{Muzykantskii_Xray}
which resolved a long-standing controversy on
orthogonality 
catastrophe in two Fermi seas\cite{Ng,Combescot,Braunecker},
little is known about
FES out of equilibrium.

Nonequilibrium 
electron states 
with structured energy distribution
were demonstrated recently\cite{Saclay_tunneling} using
diffusion-cooled nanoscale wires.
In a wire 
short enough to allow
electrons diffuse out without energy relaxation
a distribution consisting of two Fermi steps,
\be\label{eq:(1-a)n_1+an_2}
n(\epsilon)= (1-{x}) n_F(\epsilon-\mu_1)+{x} n_F(\epsilon-\mu_2)
\ee
with $\mu_{1,2}$ potentials in the leads,
was created,
imaged
using tunneling spectroscopy 
and employed to study
energy relaxation.
Similar approach\cite{split_Kondo}
was used to observe splitting 
of a Kondo resonance in a quantum dot
with a mixture of two Fermi steps injected in one of the leads.

Here we study how the FES tunneling density of states is modified 
by nonequilibrium electron distribution,
and find that it can acquire a rich and complex structure.
Since FES 
peaks at the Fermi level, one expects a multiple FES peak profile
for a multi-step distribution
of Refs.\cite{Saclay_tunneling,split_Kondo},
with each FES peak centered around a corresponding Fermi sub-level.
While the standard methods used to describe
FES in equilibrium fail, 
an exact solution can be obtained
with the help of a method proposed below
which allows to extend the FES theory to 
generic nonequilibrium systems.

Although a variety of methods is available to treat the FES problem, 
applying them outside equilibrium 
is often problematic.
The original 
approach\cite{Nozieres_deDominicis},
based on resummation of diagrammatic series,
is cumbersome and proves difficult to generalize.
Thus alternative techniques, 
most notably bosonization\cite{SchotteSchotte}, have 
been developed. The bosonization approach, however, 
relies too strongly on the 
assumption of thermodynamic equilibrium, and thus cannot be used in 
our problem.

\vspace{-4mm}
\begin{figure}[t]
\centering{
\includegraphics[width=2.5in]{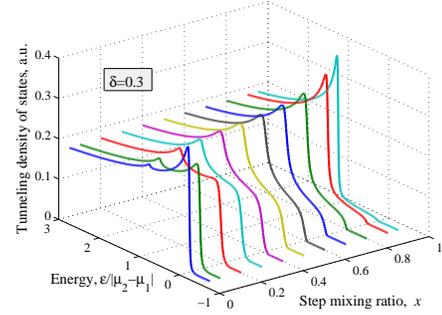}
}
\vspace{0.1cm}
\caption[]{
Fermi-edge resonance splitting (\ref{eq:ImG})
for the two-step Fermi distribution (\ref{eq:(1-a)n_1+an_2}),
with the scattering phase $\delta=0.3$.
}
\label{fig1}
\end{figure}
\vspace{-2mm}

The method used in this article 
is free of such limitation.
It allows to represent the Green's function of an 
electron tunneling in a many-body system in terms of an
appropriate functional determinant and related quantities
which are defined in a {\it one-particle} Hilbert space. 
The determinant structure accounts in an exact way 
for all the effects of Fermi statistics, as well as for the 
interaction in the final state underpinning the FES phenomenon.
Here we employ a generalization of the method of
Ref.\cite{Muzykantskii_Xray} recently used 
in nonseparable mesoscopic FES problem\cite{AbaninLevitov},
which allows to handle an arbitrary 
energy distribution. 
After developing general formalism
we focus on the two-step case (\ref{eq:(1-a)n_1+an_2}) 
and obtain a split FES profile ${\rm Im}\,G(\epsilon)$ 
in terms of the scattering phase shift $\delta$ (Fig.\ref{fig1}),
where
%
\be\label{eq:ImG}
G(\epsilon)
\propto 
\int 
\frac{1-n(\epsilon')}{
(\epsilon'-\mu_1)^{\alpha_1}(\epsilon'-\mu_2)^{\alpha_2}}
\times D(\epsilon-\epsilon')
d\epsilon'
\ee
with {\it complex exponents}
$\alpha_1=2(\delta-\tilde\delta)/\pi$,
$\alpha_2=2\tilde\delta/\pi$
and
%
\be\label{eq:delta_tilde_def}
\tilde{\delta}=
\frac{1}{2i}
\ln \lp 1-{x}+e^{2i\delta}{x}  \rp
\ee
%
The two factors in Eq.(\ref{eq:ImG})
correspond, as we will see, to the well known separation\cite{Nozieres_deDominicis}
of equilibrium FES into the `open line' and `closed loop'
contributions. The closed loop factor $D(\epsilon)$ 
equals
$\epsilon^{(\delta^2/\pi^2-1)}$ in equilibrium.
We evaluate $D(\epsilon)$ below and find that
it describes broadening of nonequilibrium FES,
with $\gamma\simeq x(1-x)|\mu_1-\mu_2|\delta^2/\pi$,
which can be attributed to
a finite effective temperature 
$T_\ast\simeq\int n(1-n)d\epsilon $.
The relation $\alpha_1+\alpha_2=2\delta/\pi$ ensures 
agreement with the equilibrium FES exponent.

Turning to the analysis, the FES Hamiltonian describes band electrons
interacting with a localized state:
\be\label{eq:H01}
\HH=\HH_0 \hat b^+\hat b + \HH_1(1-\hat b^+\hat b)
,\quad
\HH_{0,1}
=\sum_{pp'} \hat h^{(0,1)}_{pp'}a^+_p
a_{p'}
\ee
where $\hat b$, $\hat b^+$ describe 
the localized state occupation
and $\hat h^{(0,1)}_{pp'}=\epsilon_p\delta_{pp'}+V^{(0,1)}_{p-p'}$
are the single particle operators
of band electrons scattering on the charged/uncharged state
potential $V^{(0,1)}(r)$.
Tunneling from the localized state is described by
the Green's function
\be\label{eq:G_general}
G(\tau)
=\tr \lp \hat b^+(0)\hat \psi(0)\hat \psi^+(-\tau) \hat b(-\tau) \hat\rho\rp
, \quad \tau>0,
\ee
where $\hat \psi^+(\tau)=\sum_p u_p^\ast \hat a_p^+(\tau)$ 
creates an electron in the band state $\psi(r)=\sum_pu_pe^{ipr}$.
The localized state is filled prior to tunneling,
thus the density matrix
in Eq.(\ref{eq:G_general}) is
$\hat \rho = \hat \rho_e \hat b^+ \hat b$.  
Here $\hat\rho_e=\prod_p(n_pa_p^+a_p+(1-n_p)a_pa_p^+)$ 
describes band electrons
with energy distribution $n(\epsilon_p)$.
The latter quantity 
can also be written 
as an exponential of a quadratic many-body operator, 
\be\label{eq:rho_e}
\hat\rho_e = \frac1{Z} \exp\lp -\sum_p \lambda_p \hat a^+_p \hat a_p\rp
\,,\quad
e^{-\lambda_p}=\frac{n(\epsilon_p)}{1-n(\epsilon_p)}
\ee
with 
$
Z=\prod_{p} \lp 1+e^{-\lambda_p} \rp
$.
One can bring $G(\tau)$ to a standard form which depends only 
on the band electron variables by summing over the hole 
states\cite{Mahan,Nozieres_deDominicis}. This is achieved
by disentangling 
$b$ and $b^+$ from 
Eq.(\ref{eq:G_general}),
\[
\hat \psi^+(-\tau)\hat b(-\tau)=e^{-i\HH \tau}\hat \psi^+\hat b e^{i\HH \tau}
= -\hat b e^{-i\HH_1 \tau}\psi^+ e^{i\HH_0 \tau}
\]
and then using the commutation relations
\[
\hat a^+_p \hat \rho_e = e^{\lambda_p} \hat \rho_e \hat a^+_p 
\,,\quad
\hat a^+_p e^{i\HH_0\tau} = e^{-i\epsilon_p \tau} e^{i\HH_0 \tau} \hat a^+_{p}
\]
After summing over $b$, $b^+$ we obtain an expression
\be\label{eq:G_aa'}
G(\tau)
=\sum_{p,p'} u^\ast_{p'} u_p
e^{\lambda_{p'}}  e^{-i\epsilon_{p'} \tau}
\tr [ e^{-i\HH_1 \tau}e^{i\HH_0 \tau} \hat\rho_e \hat a^+_{p'} \hat a_p ]
\ee
The central point of our approach is a relation between 
the many-body operators in Eq.(\ref{eq:G_aa'})
and appropriate quantities (scattering operators and energy 
distribution) defined in a single-particle Hilbert space. 
This relation holds\cite{Klich} for any electron density matrix 
of the form of an exponential of
a quadratic many-body operator, such as 
Eq.(\ref{eq:rho_e}).

The advantage of introducing the single-particle scattering operators 
in the formalism at an early stage of the calculation
is two-fold. First, we bypass 
solution of the single particle scattering problem
which requires resummation of diagrammatic 
series\cite{Nozieres_deDominicis} for
band electron  
in the presence of a time-dependent scattering.
Second, we shall be able to construct a non-perturbative 
solution applicable to 
an arbitrary energy distribution.

We recall that Ref.\cite{Nozieres_deDominicis}
treats FES by solving the 
Dyson integral equation using a decomposition of quantities
into analytic and anti-analytic
functions of complex time variable,
made possible by breaking the 
Hilbert space into the
positive and negative frequency components. This approach
arises naturally in the equilibrium problem with a pure step 
$n(\epsilon)$
but fails for a generic distribution.
Below we develop an adequate replacement of this scheme.

The discussion in the following two paragraphs closely follows that
of Ref.\cite{AbaninLevitov}. 
First, we introduce an operator $\hat w$
defined in the single particle Hilbert space of a band electron
via the following operator product
\be\label{eq:w_def}
e^{-i\HH_1 \tau}e^{i\HH_0 \tau} \hat\rho_e = Z^{-1} \exp\lp \sum_{p,p'}
w_{p,p'}\hat a^+_p\hat a_{p'}\rp
\ee
The trace in Eq.\,(\ref{eq:G_aa'}) can be expressed
through the operator $\hat w$ as follows:
\be
\tr ( e^{-i\HH_1 \tau}e^{i\HH_0 \tau} \hat\rho_e 
\hat a^+_{p'} \hat a_p )
= 
(\hat 1+e^{-\hat w})^{-1}_{p,p'} \det ( \hat 1+e^{\hat w} ).
\ee
Our task is thus reduced to analyzing the quantity
$e^{\hat w}$ which can be expressed through 
single particle operators:
\be
e^{\hat w}=e^{-i\hat h^{(1)} \tau}e^{i\hat h^{(0)} \tau}e^{-\hat\lambda}
\ee
with the single particle 
Hamiltonian operators $\hat h^{(0,1)}$ defined in Eq.(\ref{eq:H01})
and $(e^{-\hat\lambda})_{pp'} =e^{-\lambda_p}\delta_{pp'}$.
These relations help\cite{Klich} to bring the determinant 
$\det\lp 1+e^{\hat w} \rp$ to the form
$
Z \det \lp 1- n(\epsilon) + e^{-i\hat h^{(1)} \tau}
e^{i\hat h^{(0)} \tau} n(\epsilon)\rp
$,
with the many-body effects fully accounted for by 
the algebra involved in the determinant construction. 

Next, 
the evolution operator product $e^{-i\hat h^{(1)} \tau}e^{i\hat h^{(0)} \tau}$ 
is related 
to the scattering matrix\cite{AbaninLevitov}. For one channel,
\be
\hat S \equiv e^{-i\hat h^{(1)} \tau}e^{i\hat h^{(0)} \tau}=
\delta_{t,t'}\times \cases{ e^{2i\delta},& $0<t,t'<\tau$ \cr 1, &else }
\ee
with the phase shift 
$\delta=\delta_1-\delta_0$ describing the 
effect of the resonance level changing occupancy.
This gives
\be
\det\lp 1+e^{\hat w} \rp =Z \det \lp 1+(\hat S-1) \hat n\rp
\ee
Similarly, 
$
\lp 1+e^{-\hat w} \rp^{-1}= \lp n(\epsilon) + (1- n(\epsilon))\hat S^{-1}\rp^{-1} 
n(\epsilon)
$
%
where $n$ and $S$ are operators in the Hilbert space of functions 
of time. 
The Green's function (\ref{eq:G_aa'}) then 
becomes
\bea\label{eq:G_factored}
&& G(\tau)= L(\tau) D(\tau)
,\quad
D=\det \lp 1+(\hat S-1) \hat n\rp
\\ \label{eq:L_general}
&& L=\sum_{\epsilon,\epsilon'}\tilde u^\ast_{\epsilon'}\tilde u_{\epsilon} 
e^{-i\epsilon'\tau}(1-n(\epsilon))
\lp \hat n+\hat S^{-1}(1- \hat n)\rp^{-1}_{\epsilon,\epsilon'} 
\eea
with $\tilde u_{\epsilon}=
\sum_p u_p \delta(\epsilon-\epsilon_p)$.
The factors $L$ and 
$D$ correspond, 
in the terminology of Ref.\cite{Nozieres_deDominicis},
to the open line and closed loop diagram contributions, respectively.

Once the problem is reduced to analyzing certain one-particle 
operators there are two ways to proceed. 
Given that $\hat n$ is diagonal in the energy 
domain, while $\hat S$ is diagonal in the time domain, one can choose
either representation to analyze the quantities 
in Eq.\,(\ref{eq:G_factored}). 
The former is convenient in equilibrium, 
since the $T=0$ Fermi distribution 
is just a Cauchy kernel\cite{Nozieres_deDominicis}. 
However, 
since for generic $n(\epsilon)$ the kernel
$\hat n_{t,t'}=\int n(\epsilon)e^{i\epsilon(t-t')}(d\epsilon/2\pi)$ 
is fairly complicated,
the time representation does not appear to be useful.
Here, instead, we employ the energy representation. 
We note that the operator $(\hat S-1)$ has a double step structure
$\theta(t)\theta(\tau-t)$
and argue that the contributions of the two steps 
can be treated as independent with logarithmic accuracy. 
For a single step {\it in the time domain}, the corresponding operator 
has the form of a Cauchy kernel {\it in the
energy domain}. Such energy-time duality allows to perform calculation 
in essentially the same way as in the equilibrium problem, 
with the roles of energy and time interchanged. 

Since we are primarily interested in the 
power law exponent of $G(\tau)$ rather than a prefactor,
let us consider replacing the double step
$\theta(t)\theta(\tau-t)$ by a sum of almost nonoverlapping
contributions
$\theta(t)e^{-t/\tau'}+\theta(\tau-t)e^{-(\tau-t)/\tau'}$,
$\tau'<\tau$. 
Such a replacement is reasonable since it preserves the steps at
$t=0,\tau$ and thus affects the corresponding shakeup contributions
merely by $\tau$ changed to $\tau'$ in the cutoff of the logarithms.
(In addition, we will have to adjust the extensive 
part $\ln G_{\rm lin}\propto\tau$ of the
closed loop contribution as described below.)
At the same time, since at $\tau'\lesssim\tau$ 
the two terms
do not overlap, the operator quantities
in Eq.\,(\ref{eq:G_factored}) factor into two independent
contributions. 
Employing this idea, we replace the scattering operator 
$\hat S$ by a product $\hat T_1 \hat T_2$,
where
\bea\label{eq:T_12}
&& (\hat T_1-1)_{t,t'}=\delta_{t,t'}\times \theta (t) A e^{-t/\tau'}, 
\\
&& (\hat T_2-1)_{t,t'}=\delta_{t,t'}\times \theta(\tau-t) A e^{(\tau-t)/\tau'}
\eea
($A=e^{2i\delta}-1$). 
We note that at $t\simeq 0$, where $\hat T_1$ time dependence 
has a step, 
the operator $\hat T_2$ is close to unity, while at $t\simeq\tau$,
where $\hat T_2$ has a step,
$\hat T_1$ is close unity.
This transformation allows to treat the contributions 
of $\hat T_1$, $\hat T_2$ independently, 
which greatly facilitates analysis. 

We first analyze the open line contribution (\ref{eq:L_general}).
In the $\hat S=\hat T_1\hat T_2$ approximation, 
$\tau'\lesssim\tau$, the operator in 
Eq.(\ref{eq:L_general})
is factored into independent contributions as
\be\label{eq:open_line}
\lp \hat n+\hat S^{-1}(\hat 1-\hat n)  \rp^{-1}
=\prod_{j=1,2}
\lp \hat 1+\hat B_j (\hat 1-\hat n)  \rp^{-1},
\ee
with $\hat B_j=\hat T_j^{-1}-\hat 1$.
Let us write $\hat B_1$ in the energy domain:
\be\label{eq:sigma1_def}
\hat B_1=(e^{-2i\delta}-1) \hat \sigma, \, \quad 
\hat \sigma_{\epsilon, \epsilon'}= -\frac{i}{2\pi} \frac {1}{\epsilon-\epsilon'-i/\tau'}
\ee
Hence 
$
1+\hat B_1(1-\hat n)=1-\hat \sigma+\hat \sigma f(\epsilon)
$
with $f(\epsilon)=(e^{-2i\delta}-1)(1-n(\epsilon))+1$.
To invert this operator we use analytic properties of $\hat\sigma$. 
We note that $\hat \sigma_{\epsilon, \epsilon'}$ turns into a Cauchy
kernel at large $\tau'$. Thus in this limit the operator $\hat \sigma$
projects to zero the functions $Y_+(\epsilon)$ analytic 
in the upper halfplane
of complex $\epsilon$, ${\rm Im}\, \epsilon>0$,
while $\hat \sigma^\ast=1-\hat \sigma$ 
projects to zero the functions $Y_-(\epsilon)$ 
analytic in the lower halfplane,
${\rm Im}\,\epsilon<0$. Conversely, 
$\hat \sigma^\ast$ operates
as an identity in the subspace of functions $Y_+(\epsilon)$,
while $\hat \sigma$ is an identity in the subspace of functions
$Y_-(\epsilon)$. Hence it is convenient 
to employ factorization
\be\label{eq:factoriz}
f(\epsilon)=Y_+(\epsilon) Y_-^{-1}(\epsilon)
\ee
The factors $Y_{\pm}$ are given in explicit form by 
\be\label{eq:Y_def}
\ln Y_{\pm}(\epsilon)=-\frac{1}{2\pi i} \int \frac{\ln f(\epsilon')}{\epsilon-\epsilon' \pm i0} d\epsilon'
\ee
Then, taking into account analytic properties of $Y_\pm$,
\[
\lp 1-\hat \sigma+\hat \sigma f(\epsilon)\rp^{-1}_{\epsilon, \epsilon'}=Y_-(\epsilon) 
\lp (1-\hat \sigma) Y_-^{-1}(\epsilon')+\hat \sigma Y_+^{-1}(\epsilon') \rp
\]
Similarly, the inverse $\lp \hat 1+\hat B_2(\hat 1-\hat n)\rp ^{-1}_{\epsilon, \epsilon'}$ is given by
\be\label{eq:oper2_inverted}
e^{-i\epsilon \tau} Y_+^{-1}(\epsilon)((1-\hat{\s}^{\ast} )Y_+(\epsilon')+\hat{\s}^{\ast} Y_-(\epsilon'))e^{i\epsilon'\tau}
\ee
where 
$
\hat{\s}^{\ast}_{\epsilon,\epsilon'}= (i/2\pi) (\epsilon-\epsilon'+i/\tau')^{-1}
$.
After summing over $\epsilon$, $\epsilon'$ in Eq.(\ref{eq:L_general}), 
we obtain $L(\tau)=\sum_{\epsilon}L(\epsilon)e^{-i\epsilon\tau}$
where
\be\label{eq:L_throughY}
L(\epsilon)=|u_\epsilon|^2(1-n(\epsilon))Y_+^{-2}(\epsilon)
\ee
To better understand this general result, let us consider
the two-step distribution (\ref{eq:(1-a)n_1+an_2}). 
Using Eq.\,(\ref{eq:Y_def}), we obtain
\be\label{eq:Y_doublestep}
\ln Y_+=\frac{\tilde{\delta}}{\pi} \ln \frac{\tilde\mu_2-\epsilon}{\tilde\mu_1-\epsilon}-
\frac{\delta}{\pi} \ln \frac{\xi_0-i/\tau}{\tilde\mu_1-\epsilon}  
\ee
Here  $\tilde\mu_{1,2}=\mu_{1,2}-i/\tau$ and $\tilde{\delta}$ 
is defined by Eq.(\ref{eq:delta_tilde_def}).
Substituting this result into Eq.\,(\ref{eq:L_throughY}), we obtain 
a split-peak structure with power law singularities at $\epsilon=\mu_{1,2}$:
\be\label{eq:L_final}
L(\epsilon)=
\frac{|u_\epsilon|^2 (1-n(\epsilon))(-\xi_0)^{2\delta/\pi}}{(\epsilon-\tilde\mu_1)^{2(\delta-\tilde\delta)/\pi}(\epsilon-\tilde\mu_2)^{2\tilde\delta/\pi}}
\ee
At large $|\epsilon|\gg |\mu_2-\mu_1|$, the power law form
$L\propto \epsilon^{-2\delta/\pi}$ matches the equilibrium 
result\cite{Nozieres_deDominicis}.

We now proceed to calculate the closed loop contribution (\ref{eq:G_factored}). 
First, consider a quasiclassical result, obtained by treating 
the time and energy as commuting variables: 
\be\label{eq:G_quasiclass}
\ln \det \lp 1-\hat n+\hat S\hat n\rp =
\frac{\tau}{2\pi\hbar}
\int \ln \lp 1+A n(\epsilon')\rp 
d\epsilon'
\ee
Thus we have $\det ( 1-\hat n+\hat S\hat n) = e^{-\zeta \tau}$
with complex  $\zeta=\gamma+i\epsilon_0$, where the
real part 
$\gamma$
describes broadening of the FES singularity, while  
the imaginary part $\epsilon_0$ 
describes energy offset
and can be absorbed in the phase factor $e^{-i\epsilon\tau}$. 
Evaluating the integral (\ref{eq:G_quasiclass})
for the two-step energy distribution (\ref{eq:(1-a)n_1+an_2}),
we obtain
\be\label{eq:broadening}
-\gamma = \frac{|\mu|}{4\pi\hbar} 
\ln\lp 1-4{x}(1-{x})\sin^2\delta\rp
,\quad
\mu\equiv \mu_2-\mu_1 
\ee
Thus quasiclassical FES energy structure is a broadened step. 
The power law singularity appears only beyond the quasiclassical 
approximation. To describe it one has to account the contributions 
highly nonlocal in time, corresponding to 
many low energy particle-hole excitations.

Again using the factorization approximation $S=\hat T_1\hat T_2$ with 
soft 
cutoff $e^{-t/\tau'}$, we factor the determinant (\ref{eq:L_general}) as
\be
\label{eq:closed_loop}
D=D_1D_2=\prod_{j=1,2}
\det \lp 1+(\hat T_j-1) \hat n \rp
\ee
with the two factors accounting for the contributions 
of abrupt switching at $t=0$ and $t=\tau$.
It is clear that the two determinants $D_{1,2}$ are equal, therefore, 
it is sufficient to evaluate just one of them. Let us consider 
\be\label{eq:D_def}
D_1=\det (1+(\hat T_1-1)\hat n)
\ee
The logarithm $\ln D_1$ can be represented as a sum 
\be
\ln D_1=\frac12 C_{\rm lin}+C_{\rm log}
\ee
with $C_{\rm lin}\propto \tau$ and $C_{\rm log}\propto \ln\tau$.
We have already estimated the former
(see Eq.(\ref{eq:broadening})); to obtain the latter,
we consider variation $\Delta \ln D_1$ caused by a change 
in the distribution
$n(\epsilon)$. 
With the help
of the formula $\Delta \ln \det U=\tr (U^{-1}\Delta U)$, we obtain
\be\label{eq:det_differentiated}
\Delta \ln D_1 = 
A\, {\rm tr} \lb \lp 1-\hat{\s}^{\ast}+\hat{\s}^{\ast}(A\hat n+1) \rp^{-1}
\hat{\s}^{\ast} \Delta \hat n \rb
\ee
To evaluate this expression we employ factorization 
\be\label{eq:A_factorized}
An(\epsilon)+1=X_-(\epsilon)X_+^{-1}(\epsilon)
\ee
to rewrite the right hand side of Eq.\,(\ref{eq:det_differentiated}) as
\[
{\rm tr} \lp X_+(\epsilon)\lb(1-\hat{\s}^{\ast})X_+^{-1}(\epsilon')+\hat{\s}^{\ast}X_-^{-1}(\epsilon')\rb
\hat{\s}^{\ast} A\,\Delta n(\epsilon')
\rp
\] 
Since $X_+(\epsilon)$ is analytic in the upper halfplane, 
$\int (1-\hat{\s}^{\ast})_{\epsilon,\epsilon'}X_+^{-1}(\epsilon')\hat{\s}^{\ast}_{\epsilon',\epsilon''}d\epsilon' =0$ and
Eq.\,(\ref{eq:det_differentiated}) becomes
\be\label{eq:det_simplified}
\Delta \ln D_1=
A\, {\rm tr}\lp X_+ \hat{\s}^{\ast} X_-^{-1}
\Delta\hat n
\rp 
\ee
From Eq.(\ref{eq:det_simplified}) we extract the contribution corresponding to $C_{\rm log}$:
\be\label{eq:log_contr}
\Delta C_{\rm log}=
\frac{i}{2\pi} \int \frac{A\Delta n(\epsilon)}{1+An(\epsilon)} \frac{\p}{\p\epsilon} \ln X_+(\epsilon)   \,d\epsilon 
\ee
So far the particular form of the distribution function 
$n(\epsilon)$ did not matter,
since Eqs.(\ref{eq:L_throughY}),(\ref{eq:log_contr}) 
solve the problem for arbitrary $n(\epsilon)$.
For the two-step distribution (\ref{eq:(1-a)n_1+an_2}) 
the function $\ln X_+(\epsilon)$ takes the following form
\be\label{eq:X_doublestep}
\ln X_+(\epsilon)=-\frac{\delta}{\pi}\ln \frac{\mu_1-\epsilon-i/\tau}{-\xi_0-i/\tau}-
\frac{\tilde{\delta}}{\pi}\ln \frac{\mu_2-\epsilon-i/\tau}{\mu_1-\epsilon-i/\tau}
\ee
with $\xi_0\simeq E_F$ an ultraviolet cutoff.
To find $\Delta C_{\rm log}$ we consider variation of 
(\ref{eq:(1-a)n_1+an_2}) with respect to ${x}$, 
\[
\Delta n(\epsilon)=\Delta {x} \times \cases{ 1,& $\mu_1<\epsilon<\mu_2$ \cr 0, &else }
\]
Eq.(\ref{eq:log_contr}) then yields a relation
\be\label{eq:log_doublestep}
\frac{\p C_{\rm log}}{\p{x}} = 
iA \frac{\tilde{\delta}\ln (1+\mu^2\tau^2)-\delta \ln (1-i\mu\tau)}{2\pi^2 (1+A{x})} 
\ee
Using the known value of $C_{\rm log}$ for the equilibrium distribution, 
we integrate Eq.\,(\ref{eq:log_doublestep}) over $0<{x}'<{x}$
and obtain the logarithmic term $C_{\rm log}$ as a function of ${x}$, 
\be\label{eq:Clog_final}
e^{C_{\rm log}({x},\tau)}=
\frac{(1-i\mu\tau)^{\delta\tilde{\delta}/\pi^2}}{(1+\mu^2\tau^2)^{\tilde{\delta}^2/2\pi^2} }
(-i\tau\xi_0)^{-\delta^2/2\pi^2} 
\ee
Finally, restoring the exponential from Eqs.\,(\ref{eq:G_quasiclass}),(\ref{eq:broadening}), 
which is responsible for broadening of the FES, we obtain the closed loop factor
$D(\tau)=e^{C_{\rm log}({x},\tau)}\exp(-\gamma\tau)$. 

The function $D(\epsilon)=\int e^{i\epsilon \tau} D(\tau) d\tau$ defines broadening of
singularities in the split FES, Eq.(\ref{eq:ImG}). We note that the broadening 
is relatively insignificant at $\delta\ll 1$, since 
$\alpha_{1,2}\propto \delta$ while
both $\gamma$ and the 
exponents in (\ref{eq:Clog_final}) are of order $\delta^2$ at
small $\delta$. Thus, although FES broadening is present for a split Fermi step,
its magnitude is not large enough to smear the split peak FES profile.


This work has benefited from the discussions 
with Boris Muzykantskii 
and was  
supported by MRSEC Program of
the National Science Foundation
(DMR 02-13282).

\end{multicols}


\vspace{-5mm}
\begin{thebibliography}{99}
\vspace{-5mm}

\bibitem{review}
\emph{Electron Transport in Quantum Dots}, 
NATO ASI Conference Proceedings, 
ed. by L.\,P.\,Kouwenhoven, G.\,Schon, L.\,L.\,Sohn (Kluwer, Dordrecht, 1997)

\bibitem{Kaminski}
A.\,Kaminski, Yu.\,V.\,Nazarov, and L.\,I.\,Glazman
\emph{Phys. Rev.} B {\bf 62}, 8154 (2000)

\bibitem{Moore}
J.\,E.\,Moore and X.-G.\,Wen
\emph{Phys. Rev. Lett.} {\bf 85}, 1722 (2000)

\bibitem{Coleman}
P.\,Coleman, C.\,Hooley, and O.\,Parcollet
\emph{Phys. Rev. Lett.} {\bf 86}, 4088 (2001)

\bibitem{Konik0102}
R.\,M.\,Konik, H.\,Saleur, and A.\,Ludwig,
\emph{Phys. Rev. Lett.} {\bf 87}, 236801 (2001);
\emph{Phys. Rev.} B {\bf 66}, 125304 (2002)

\bibitem{Paaske}
J.\,Paaske, A.\,Rosch, and P.\,W\"{o}lfle,
\emph{Phys. Rev.} B {\bf 69}, 155330 (2004)

\bibitem{Komnik04}
A.\,Komnik and A.\,O.\,Gogolin,
\emph{Phys. Rev.} B {\bf 69}, 153102 (2004)

\bibitem{Mahan} 
G.\,D. Mahan, 
\emph{Phys. Rev.} {\bf 163}, 612 (1967)

\bibitem{Nozieres_deDominicis} P.\,Nozi\`eres and C.\,T.\,De Dominicis,
{\it Phys. Rev.} {\bf 178}, 1097 (1969)

\bibitem{MatveevLarkin} K.\,A.\,Matveev and A.\,I.\,Larkin,
{\it Phys. Rev. B}{\bf 46}, 15337 (1992)

\bibitem{Geim}
A.\,K. Geim, \emph{et al.},
\emph{Phys. Rev. Lett.} {\bf 72}, 2061 (1994)

\bibitem{Ogawa92}
T. Ogawa, A. Furusaki and N. Nagosa, 
\emph{Phys. Rev. Lett.} {\bf 68}, 3638 (1992)

\bibitem{Prokof'ev}
N. V. Prokof'ev, 
\emph{Phys. Rev.} B {\bf 49}, 2148 (1994)

\bibitem{Komnik97}
A. Komnik, R. Egger, and A. O. Gogolin,
\emph{Phys. Rev.} B {\bf 56}, 1153 (1997) 

\bibitem{Affleck94}
I. Affleck and A. Ludwig, 
\emph{J. Phys.} A {\bf 27}, 5375 (1994)

\bibitem{Bascones}
E. Bascones, C. P. Herrero, F. Guinea, and G. Sch\"on,
\emph{Phys. Rev.} B{\bf 61}, 16778 (2000)

\bibitem{Muzykantskii_Xray} 
B. Muzykantskii, N. d'Ambrumenil and B. Braunecker,
\emph{Phys. Rev. Lett.} {\bf 91}, 266602 (2003);
N. d'Ambrumenil, B. Muzykantskii, cond-mat/0405475
 
\bibitem{Ng}
Tai-Kai Ng,
\emph{Phys. Rev.} B {\bf 54}, 5814 (1996)


\bibitem{Combescot}
M. Combescot and B. Roulet, 
\emph{Phys. Rev.} B {\bf 61}, 7609 (2000)

\bibitem{Braunecker}
B. Braunecker,
\emph{Phys. Rev.} B {\bf 68}, 153104 (2003)

\bibitem{Saclay_tunneling}
H. Pothier, \emph{et al.},
\emph{Phys. Rev. Lett.} {\bf 79}, 3490 (1997)

\bibitem{split_Kondo} 
S. De Franceschi, \emph{et al.}
\emph{Phys. Rev. Lett.} {\bf 89}, 156801 (2002)



\bibitem{SchotteSchotte}
K.\,D. Schotte and U. Schotte, 
\emph{Phys. Rev.} {\bf 182}, 479 (1969)

\bibitem{AbaninLevitov}
D. A. Abanin and L. S. Levitov, 
\emph{Phys. Rev. Lett.} {\bf 93}, 126802 (2004)


\bibitem{Klich}
I. Klich, 
in \emph{Quantum Noise in Mesoscopic Systems},
ed. by Yu. V. Nazarov (Kluwer, Dordrecht, 2003);
cond-mat/0209642
 


\end{thebibliography}
\end{document}